\documentstyle[psfig,12pt]{l-aa}

\newcommand{\apj}[2]{ApJ #1, #2}
\newcommand{\aps}[2]{ApJS #1, #2} 
\newcommand{\am}[2]{A\&A #1, #2}
\newcommand{\as}[2]{A\&AS #1, #2} 
\newcommand{\aj}[2]{AJ #1, #2}
\newcommand{\ara}[2]{ARA\&A #1, #2} 
\newcommand{\asp}[2]{Ap\&SS #1, #2} 
\newcommand{\mn}[2]{MNRAS #1, #2}
\newcommand{\qj}[2]{QJRAS #1, #2}  
\newcommand{\cmtwo}{cm$^{-2}$}
\newcommand{\cmthree}{cm$^{-3}$}

\newcommand{\um}{$\mu$m}                                 
\newcommand{\lsol}{$L_{\odot}$}      
 
\newcommand{\msol}{$M_{\odot}$}
\newcommand{\rsol}{$R_{\odot}$}
\newcommand{\mdot}{\dot{M}}
\newcommand{\msunyr}{$M_{\odot}$ yr$^{-1}$}

\newcommand{\lsim}{\;\lower.6ex\hbox{$\sim$}\kern-7.75pt\raise.65ex\hbox{$<$}\;}
\newcommand{\gsim}{\;\lower.6ex\hbox{$\sim$}\kern-7.75pt\raise.65ex\hbox{$>$}\;}

\newcommand{\amin}{$^{\prime}$}                   
\newcommand{\asec}{$^{\prime \prime}$}

\newcommand{\hi}{{\it High}}
\newcommand{\lo}{{\it Low}}
\newcommand{\be}{\begin{equation}}
\newcommand{\ee}{\end{equation}}
\newcommand{\hii}{H{\sc ii}}
\begin{document}
%

%
  \thesaurus{06(08.03.4, 08.06.2, 08.16.5, 09.08.1, 13.20.1) }
  \title{A search for precursors of Ultracompact H{\sc ii} Regions in a sample of 
luminous IRAS sources.}
\subtitle{III: Circumstellar Dust Properties}
  \author{ S. Molinari\inst{1,}\inst{2,}\inst{3}  
  \and J. Brand\inst{2}
  \and R. Cesaroni\inst{4}
  \and F. Palla\inst{4}}
%
     \offprints{ S. Molinari, IPAC/Caltech, molinari@ipac.caltech.edu}
%
%
%
  \institute{Infrared Processing and Analysis Center, California Institute of 
Technology, MS 100-22, Pasadena, CA 91125, USA
\and Istituto di Radioastronomia-CNR, Via Gobetti 101, I-40129
  Bologna, Italy
\and Universit\`a degli Studi di Bologna, Dipartimento di Astronomia, Via Ranzani 1, I-40127 Bologna, Italy
\and Osservatorio Astrofisico di Arcetri, Largo E. Fermi 5, I-50125 Firenze, Italy 
}
%
%
\date{Received 13 September 1999/ Accepted date}
%
%
\maketitle
\markboth{Molinari et al.: A search for precursors of UC\hii. III}{}

\begin{abstract}
The James Clerk Maxwell Telescope has been used to obtain submillimeter
and millimeter continuum photometry of a sample of 30 IRAS sources
previously studied in molecular lines and centimeter radio continuum.
All the sources have IRAS colours typical of very young stellar objects
(YSOs) and are associated with dense gas. In spite of their high
luminosities (L$\gsim 10^4$~\lsol),  only ten of these sources are also
associated with a  radio counterpart. In 17 cases we could identify a
clear peak of millimeter emission associated with the IRAS source, while
in 9 sources the millimeter emission was either extended or faint and a
clear peak could not be identified; upper limits were found in 4 cases
only. 

The submm/mm observations allow us to make a more accurate estimate  of
the source luminosities, typically of the order of 10$^4$~\lsol. Using
simple greybody fitting model to the observed spectral energy
distribution, we derive global properties of the circumstellar dust
associated with the detected sources. We find that the  dust temperature
varies from 24~K to 45~K, while the exponent of the dust emissivity 
$vs$ frequency power-law spans a range 1.56$<\beta <$2.38, characteristic
of  silicate dust; total circumstellar masses range up to more than
500~\msol.

We present a detailed analysis of the sources associated with millimeter
peaks, but without radio emission. In particular, we find  that for
sources with  comparable luminosities, the total column densities
derived from the dust  masses do not distinguish between sources with
and without radio counterpart. We interpret this result as an indication
that dust does not play a dominant role in inhibiting the formation of
the \hii\ region. We examine several scenarios for their origin in terms
of newborn ZAMS stars and although most of these (e.g. optically thick
\hii\ regions, dust extinction of Lyman photons, clusters instead of
single sources) fail to explain  the observations, we cannot exclude
that these sources are young stars already on the ZAMS with modest
residual accretion that quenches the expansion of the \hii\ region, thus
explaining the lack of radio emission in these bright sources.  Finally,
we consider the possibility that the IRAS sources are  high-mass
pre-ZAMS (or pre-H-burning) objects  deriving most of the emitted
luminosity from accretion.

\end{abstract}

\section{Introduction}

The last few years have seen a rapidly growing observational activity
aimed at the identification of intermediate- and high-mass star forming
sites exhibiting a wide range of evolutionary stages from UltraCompact
\hii\  (UC\,\hii) regions (Wood \& Churchwell~\cite{WC89}), to ``Hot
Cores'' (Cesaroni et al.~\cite{Cetal94}), and proto-Ae/Be stars
(Hunter et al.~\cite{Hetal98}; Molinari et al.~\cite{Metal98b}).  The
characterization of the earliest stages of high-mass star formation,
given  their shorter evolutionary timescales, is more difficult than for
low-mass  objects. The observational approach to the search of the
youngest high-mass forming objects was first formulated by Habing \&
Israel (\cite{HI79}): the likely candidates must have high luminosity,
be embedded in dense circumstellar environments, and they should not be
associated with \hii\ regions.

We have undertaken a systematic study aimed at the identification of a
sample of massive protostellar candidates; the whole process is
summarized in Fig.~\ref{selection}. Initially, a list of 260 sources
with 60\um\ flux greater than 100~Jy was compiled from the IRAS-PSC2,
according to the colour criteria of Richards et al. (\cite{Retal87}) for
compact molecular clouds. This sample was then divided into two groups
according to their [25$-$12] and [60$-$12] colours: the \hi\ sources,
which have [25$-$12]$\geq$0.57 and [60$-$12]$\geq$1.3 characteristic of
association with UC\,\hii\ regions (Wood \& Churchwell~\cite{WC89}), and
the \lo\ sources, with [25$-$12]$<$0.57 or [60$-$12]$\leq$1.3. We note
that the [25$-$12] and [60$-$12] IRAS colours of \lo\ sources are
different also from those of T Tau or Herbig Ae/Be stars, but are
similar to those of normal \hii\ regions (Palla et al.~\cite{Petal91}).
The lower H$_2$O maser detection rate found  towards \lo\ sources (a
factor of 3 lower than for \hi\ sources) was interpreted  as an
indication of relative youth, and we concluded that the \lo\ group might
contain a fraction of young sources whose formation process has not yet
proceeded far enough to produce a fully-developed ZAMS star  (Palla et
al.~\cite{Petal91}); the \lo\ group thus represents an optimum  target
group to search for high-mass protostars.

\begin{figure}
\centerline{\psfig{figure=h1750f1.0,height=14cm,angle=0}}
\caption[]{Flow-chart for the selection of the sample of 
intermediate-to-high mass candidate protostellar objects. The final
sample consists of 12 objects detected in the (sub-)mm and not
associated with radio continuum. However, one of these objects (\#12 in
Table~2) has been recently detected  at 3.6 cm, as explained in
footnote~\ref{newvla}.)}
\label{selection}
\end{figure}

We observed the NH$_3$(1,1) and (2,2) lines in a subsample of 80 \hi\
and 83 \lo\ sources to check for association with dense gas (Molinari
et  al.~\cite{Metal96}, hereafter Paper~I).  A result was that the
linewidth ratio $\Delta v_{22}/\Delta v_{11}$ is correlated with the
[25$-$12] colour, increasing from values $\lsim$1 to values $\gsim$1
going from \lo\ to \hi\ sources. We speculated that lower linewidth
ratios in \lo\ sources were indicative of a lesser degree of activity in
the central regions with respect to the \hi\ sources.  A critical test
to our conjecture was to verify the occurrence  of  radio continuum
emission from \hi\ and \lo\ sources.  Molinari et al. (\cite{Metal98a},
hereafter Paper~II)  observed with the VLA at 2 and 6-cm, 37 \lo\ and 30
\hi\  sources  with ammonia detections. We found that 76\% of \lo\ and 
57\% of \hi\ sources are not associated with UC\,\hii\ regions 
\footnote{\label{newvla}Recent VLA  (3.6 cm - D config.) observations 
indicate that two \lo\ sources  (\#3 and \#12 in Table~2) may be
associated with a radio counterpart.  The detected  signals are
compatible with the non-detections at 2 and 6 cm  reported for these two
sources in Paper II.}, confirming the goodness of the FIR colour-based
separation between \hi\ and \lo\ as an indicator of presence/absence of
a compact radio counterpart, and further reinforcing our assumption
about the relative youth of the \lo\ group; this idea is also supported
by the recent identification (Molinari et al.~\cite{Metal98b}) of a
massive  Class 0 (Andr\'e et al.~\cite{AWTB93}) object in the \lo\
group.

The purpose of the present observations is twofold. On the one hand we
wish to complete the selection process started by Palla et al.
(\cite{Petal91}) by identifying those objects of the initial sample
which are associated with peaks of dense gas and dust and do not show a
radio continuum counterpart (see Fig.~\ref{selection} and
Sect.~\ref{statistics}). On the other hand, in order to verify  that
different evolutionary stages are present in the \lo\ group, we need to
understand the physical nature of the distinction between \lo\ sources
with and without radio counterpart: is the \hii\ region really absent in
the latter group, or is some mechanism, independent of the evolutionary
state of the sources, responsible for inhibiting the formation of the
\hii\ region ? It is a general result from radio continuum observations
of UC\,\hii\ regions (e.g. Wood \& Churchwell~\cite{WC89}; Paper~II)
that the Lyman continuum flux required to explain the observed radio
continuum emission is generally lower than what is expected based on the
luminosity and spectral type of the ionizing star. Dust certainly plays
a role by absorbing a relevant fraction of the ionizing UV continuum 
(Aannestad~\cite{A89}) and one may ask whether the high rate of non
detection in radio continuum for the \lo\ sources might not be accounted
for by the properties of the dust in their circumstellar environment.
Millimeter continuum observations are therefore mandatory to shed light
on this issue. So far, the only information regarding the association of
the \lo\ sources with dense circumstellar environments comes from the
single-pointing ammonia measurements (Paper~I) and it is important to
check for the presence of dense and compact cores. The present paper
describes such observations; details of the observations and data
reduction procedures can be  found in Sect.~\ref{obs}, while data
analysis methods and results are described in Sect.~\ref{results}. The
derived  global properties and the nature of the detected sources are
discussed in  Sect.~\ref{global} and \ref{lownature};  the main
conclusions are summarized in  Sect.~\ref{conclusions}.

\section{Observations}
\label{obs}
Observations were performed with the James Clerk Maxwell Telescope
(JCMT) from 3 to 5 September 1994, and the UKSERV program provided
service observations on several occasions between November 1994 and June
1995. We observed 30 \lo\ sources, 10 of which are associated with
radio  continuum (Paper~II, but see footnote~\ref{newvla}). In each
observing session the common user UKT14  bolometer (Duncan et
al.~\cite{Detal90}) was used, with a focal plane aperture of 65~mm at
all wavelengths; this corresponds to a $\sim 18$\asec.5 HPBW for
wavelengths from 0.35 to 1.1 mm, increasing to 19\asec.5 and 27\asec at
1.3 and 2.0 mm respectively (Sandell~\cite{S94}). Azimuthal chopping was
done with an amplitude of 60\asec, and frequency of 7.813~Hz. 

Observations were centered on the IRAS PSC-2 coordinates and one or more
cross maps in [Az, El] at 1.1 mm with 10\asec\ spacing (called a
FIVEPOINTS cycle) were done to maximize the signal and locate the
position of the millimeter emission peak, where subsequently
observations  in the other bands were made. The control computer
automatically estimated the centroid position from each FIVEPOINTS
cycle; if this position was more distant than 3-4\asec\ from the center
of the cross map, another FIVEPOINTS cycle centered on this new position
was performed. Obviously no photometry was done when no 1.1~mm emission
was detected during the maximization procedure, or when the emission was
faint and diffuse and an emission peak could not identified. During the
September 1994 run only 0.8--2.0~mm photometry could be done as 
weather conditions prohibited observations at shorter wavelengths. In
these cases UKSERV provided the needed photometry, pointing at the
previously determined 1.1~mm peak and performing FIVEPOINTS cycles at
0.45~mm to estimate possible shifts between emission centroids at
different wavelengths. UKSERV also provided complete 0.35--2.0~mm
photometry of the few sources which had not been observed in the
September 1994 run. 
\begin{table}[h]
\begin{flushleft}
\caption{Journal of Observations}
\label{jou}
\begin{tabular}{lcccccl}\\ \hline
Date/code & $\overline{\tau _{CSO}}$ & Band & $\tau$ & Gain & Int. & \\ 
dd-mm-yy & & (mm) & & (Jy/mV) & unc. & \\ \hline 
03-09-94/A & 0.28 & 0.8 & 1.42 & 17.31 & 6\% &  \\
   	   &	& 1.1 & 0.58 & 12.82 & 3\% &  \\
	   &	& 1.3 & 0.48 & 12.70 & 2\% &  \\
	   &	& 2.0 & 0.23 & 36.98 & 1\% &  \\ \hline
04-09-94/B & 0.22 & 0.8 & 1.67 & 12.07 & 11\% & 1 \\
	   &	& 1.1 & 0.66 & 11.87 & 5\% & 1 \\
	   &	& 1.3 & 0.54 & 11.74 & 3\% & 1 \\
	   &	& 2.0 & 0.29 & 34.25 & 2\% & 1 \\ \hline
05-09-94/C & 0.15 & 0.8 & 1.16 & 10.61 & 5\% & \\
	   &	& 1.1 & 0.39 & 12.23 & 3\% & \\
	   &	& 1.3 & 0.19 & 14.52 & 2\% & \\
	   &	& 2.0 & 0.26 & 31.32 & 1\% & \\ \hline
30-11-94/S-D & 0.02 & 0.45& $-$  & 24.09 & 10\% & 2 \\ 
           &        & 0.8 & $-$  & 8.12  & 5\% & 2 \\
	   &      & 1.3 & $-$  & 10.98 & 15\% & 2 \\ \hline
19-01-95/S-E & $-$  & 0.35 & 1.09 & 19.70 & 11\% & \\
	   &	& 0.45 & 1.02 & 14.64 & 9\% & \\ \hline
20-01-95/S-F & $-$  & 0.35 & $-$ & 57.74 & 10\% & 3 \\ 
	   &	& 0.45 & $-$ & 35.82 & 10\% & 3 \\
	   &	& 0.8  & $-$ & 9.43  & 5\%  & 3 \\
	   &	& 1.1  & $-$ & 10.69 & 4\%  & 3 \\
	   &	& 1.3  & $-$ & 14.13 & 15\% & 3 \\
	   &	& 2.0  & $-$ & 37.07 & 6\% & 3 \\ \hline
06-06-95/S-G & $-$  & 0.35 & 0.77 & 65.37 & 5\% & \\
	   &	& 0.45 & 1.02 & 19.12 & 9\% & \\ \hline
\end{tabular}\\
\parbox{3.5in}{\it Notes to Table:}\\
\parbox{3.5in}{1 - Airmass coverage of the standards was not sufficient
to determine reliable optical depths and gains. So, standards
done in this night were merged with the same standards done in the
previous night, as the signal from the same standards at same airmass in
the two nights is comparable within few \%.}\\
\parbox{3.5in}{2 - Only 1 source and 1 standard were observed in this
night.}\\
\parbox{3.5in}{3 - Airmass coverage of the observed standards was not
sufficient to determine optical depths and gains from Eq.~(\ref{red2}).
Each observed source was reduced using a standard observed immediately
before or after and at comparable (within 0.1) airmasses.}\\
\end{flushleft}
\end{table}

Standard sources from the compilation of Sandell (\cite{S94}) were
observed at different airmasses to allow an estimate of the optical
depths $\tau_{\lambda}$ and detector gains $G_{\lambda}$ (in Jy/mV) in
the various bands. Following Stevens \& Robson (\cite{SR94}), the flux
$F_{\lambda}$ (in Jy) and  the signal $S_{\lambda}$ (as measured in mV
from the detector) of a source at a given airmass A can be expressed as

\be
\ln~F~-~\ln~S~=~\ln~G_{\lambda}~+~\tau_{\lambda}~A.
\label{red2}
\ee

We thus have a set of values [$\ln~F - \ln~S$, $A$] for each standard
(where $S$ and $A$ come from the observations, and $F$ is tabulated by
Sandell~\cite{S94}); a linear regression using all standards observed in
the course of a night will then provide the $\tau_{\lambda}$ (the slope)
and $G_{\lambda}$ (e-base power of the intercept). These values are then
used to compute the standards' fluxes which are compared to the tabulated
values (Sandell ~\cite{S94}) to get an estimate of the standards'
intercalibration uncertainties in all bands: these are added in
quadrature with the intrinsic statistical uncertainties of the
measurements to get the total uncertainty. 

Information about the  observations is summarized in Table~\ref{jou}
where  we give: [Column~1] date of observations and a code used to
identify the session, where `S' stands for `service observations';
[Column~2] the  1.3~mm optical depth as measured through continuous
skydips by the nearby Caltech Submillimeter Observatory (CSO); [Column~3]
the central wavelength of the photometric  band; [Columns~4-5] optical
depth and detector gain in each band; [Column~6] intercalibration
uncertainty; [Column~7] notes.

If the airmass coverage of the standards was not sufficient to derive
reliable optical depths and gains using Eq.~(\ref{red2})(sessions B and
S-F), or if just one standard was observed (session S-D), each source was
reduced using a standard at comparable airmass and observed immediately
before or after the source; in this case only the gain is reported in
Table~\ref{jou}, and the intercalibration uncertainty was conservatively
assumed to be equal to the maximum, for each band, of the uncertainties
estimated for all the other nights. 

\begin{figure*}
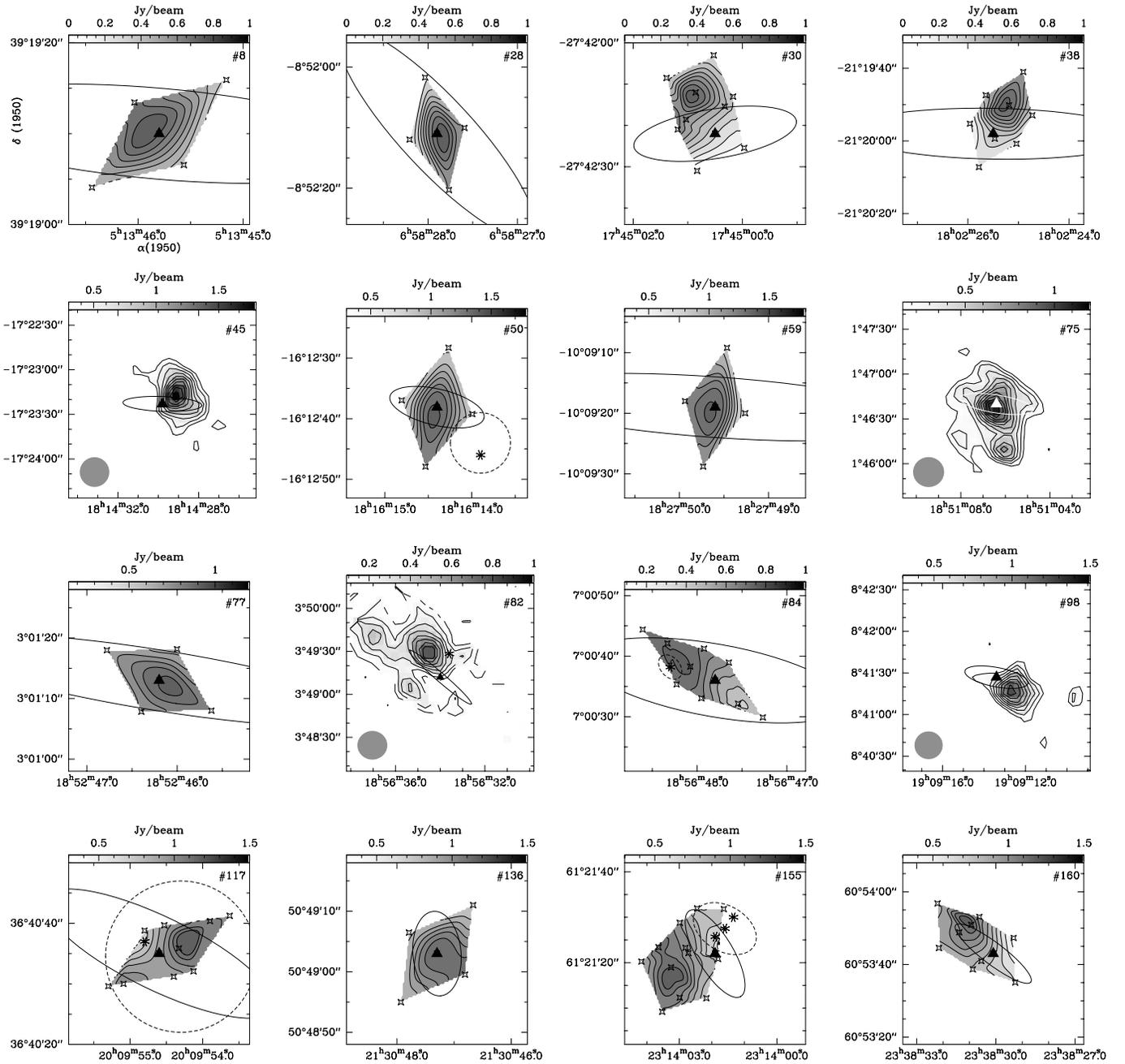

\centerline{\hbox{
\psfig{figure=h1750f2.1,height=2.5in}
\psfig{figure=h1750f2.2,height=2.5in}
\psfig{figure=h1750f2.3,height=2.5in}
\psfig{figure=h1750f2.4,height=2.5in}
}}
\vspace{-.75in}
\centerline{\hbox{
\psfig{figure=h1750f2.5,height=2.5in}
\psfig{figure=h1750f2.6,height=2.5in}
\psfig{figure=h1750f2.7,height=2.5in}
\psfig{figure=h1750f2.8,height=2.5in}
}}
\vspace{-.75in}
\centerline{\hbox{
\psfig{figure=h1750f2.9,height=2.5in}
\psfig{figure=h1750f2.10,height=2.5in}
\psfig{figure=h1750f2.11,height=2.5in}
\psfig{figure=h1750f2.12,height=2.5in}
}}
\vspace{-.75in}
\centerline{\hbox{
\psfig{figure=h1750f2.13,height=2.5in}
\psfig{figure=h1750f2.14,height=2.5in}
\psfig{figure=h1750f2.15,height=2.5in}
\psfig{figure=h1750f2.16,height=2.5in}
}}
\caption{Maps of the $\lambda=1.1$mm emission for the \lo\ sources where
a  clear emission peak was found. Most of them are reconstructed from
the FIVEPOINTS crosses made to maximize the source signal and locate the
peak. On-the-fly maps are presented for sources \#45, 75, 82 and 98 (the
grey circle on the bottom left indicates the telescope beam FWHM);
note  that the reconstruction of the maps with the NOD2 package was only
possible at the telescope. It was not possible to make a very careful
removal of mapping artifacts, and features other than the main peak are
to be considered uncertain. The full triangles represent the nominal
position of the IRAS source; the associated 1$\sigma$ error ellipse is
also  indicated with a full line. The small empty star-like symbols
represent the observed positions during the FIVEPOINTS cycles. The
positions and extensions (when resolved) of the associated radio
counterparts are indicated as asterisks and dashed ellipses
respectively. Contours levels are spaced at 10\% of the maximum flux,
and a greyscale is also superimposed for clarity. 16 out of 17 maps are
shown because no FIVEPOINTS cycles for source \#12 was done; new
unpublished SCUBA  maps show that the IRAS position for \#12, used for
the pointing in the present  observations, coincides with a millimeter
continuum peak.}
\label{maps}
\end{figure*}

Using information in Table~\ref{jou} we can compute  fluxes of target
sources in all bands. Four sources (\#45, 75, 82 and 98) were mapped {\it
on-the-fly} at 1.1~mm by chopping in an adjacent field with the
telescope stepping by 4\asec\ and integrating 2 seconds on each point;
the final maps were reconstructed using the NOD2 software, and cover a
2\amin $\times$ 2\amin\ field approximately. For other sources, limited
information about the spatial distribution of the emission can be
derived from the  FIVEPOINTS cycles done to locate the emission peak.
All the maps are presented in Fig.~\ref{maps}.

\section{Results and data analysis}
\label{results}
\begin{table*}
\begin{flushleft}
\caption{Observations}
\label{obstab}
\tabcolsep0.15cm
\begin{tabular}{lccccrrrrrr}\hline
(1) & (2) & (3) & (4) & (5) & \multicolumn{1}{c}{(7)} &
\multicolumn{1}{c}{(7)} & \multicolumn{1}{c}{(8)} & \multicolumn{1}{c}{(9)} & 
\multicolumn{1}{c}{(10)} & \multicolumn{1}{c}{(11)} \\
Mol \#$^{\clubsuit}$ & Obs. code & \multicolumn{2}{c}{IRAS Pos.} & 
$\Delta_{1.1mm Pos.}$ & \multicolumn{6}{c}{Observed Fluxes (Jy)} \\ 
\cline{3-4}\cline{5-5}\cline{6-11}
 & & $\alpha$(1950) & $\delta$(1950) &  & \multicolumn{1}{c}{0.35} & 
\multicolumn{1}{c}{0.45} 
 & \multicolumn{1}{c}{0.8} & \multicolumn{1}{c}{1.1} & 
 \multicolumn{1}{c}{1.3} & \multicolumn{1}{c}{2.0} \\ 
 \hline\hline
\multicolumn{11}{c}{Detected Sources} \\ \hline
8 & S-F & 05:13:45.8 & +39:19:09.7 & 0,0 & 17(2) & 7.0(0.7) & 1.05(0.02) & 
0.38(0.02) & 0.29(0.06) & 0.09(0.05) \\
\underline {\bf (12)$^a$} & S-F & 05:37:21.3 & +23:49:22.0 & 0$^b$, 0$^b$ &  41(5) 
& 18(2) & 2.73(0.03) & 0.91(0.04) & 0.7(0.1) & 0.3(0.1) \\
28 & S-F & 06:58:27.9 & $-$08:52:11.0 & 0, 0 &  5(1) & 
2.2(0.4) & 0.36(0.03) & 0.120(0.03) & 0.13(0.05) & 0.16(0.08) \\
30 & S-E/B$^{\rm c}$ & 17:45:00.5 & $-$27:42:22.0 & +6, +10 & 27(3) & 
15(2) & 1.2(0.3) & 0.40(0.07) & 0.21(0.08) & $<$0.06 \\
38 & S-E/B$^{\rm c,d}$ & 18:02:25.5 & $-$21:19:58.0 & $-$3, +8 & 80(9) & 41(4) & 
3.9(0.5) & 1.27(0.08) & 0.89(0.04) & 0.24(0.06) \\
45 & S-E/B$^{\rm c}$ & 18:14:29.8 & $-$17:23:23.0 & $-$9, +4 & 120(10) & 60(5) & 
4.0(0.5) & 1.63(0.09) & 1.20(0.04) & 0.29(0.06) \\
\underline {\bf 50} & B & 18:16:14.4 & $-$16:12:38.1 & 0, 0 & 24(3) & 
14(1) & 1.9(0.3) & 0.44(0.04) & 0.29(0.03) & 0.16(0.08) \\
59 & S-E/C$^{\rm c}$ & 18:27:49.6 & $-$10:09:19.0 & 0, 0 & 18.6(0.2) & 
9.3(0.9) & 1.5(0.1) & 0.33(0.03) & 0.33(0.03) & 0.18(0.04) \\
75 & S-G/A & 18:51:06.4 & +01:46:40.0 & 0, 0 & 43(5) & 24(2) & 
3.4(0.3) & 1.16(0.05) & 0.75(0.05) & 0.14(0.08) \\
77 & S-G/A & 18:52:46.2 & +03:01:13.0 & $-$2, $-$2 &  26(2) & 12(1) & 
1.5(0.2) & 0.56(0.04) & 0.39(0.04) & 0.20(0.06) \\
\underline {\bf 82} & S-G/C & 18:56:34.0 & +03:49:12.1 & +8, +18 & 
43(3) & 23(2) & 2.8(0.2) & 0.87(0.03) & 0.52(0.05) & 0.18(0.08) \\
\underline {\bf 84} & S-G/B & 18:56:47.8 & +07:00:36.1 & +6, +3 & 18(1) & 8.0(0.8) 
& 1.2(0.2) & 0.45(0.04) & 0.32(0.02) & $<$0.4 \\
98 & S-G/B & 19:09:13.4 & +08:41:27.1 & $-$10, $-$10 & 68(4) & 
32(3) & 5.8(0.7) & 1.6(0.1) & 1.09(0.06) & 0.4(0.1) \\
\underline {\bf 117} & A & 20:09:54.6 & +36:40:35.1 & $-$5, +2 & & & 
0.9(0.2) & 0.25(0.04) & 0.20(0.04) & $<$0.4 \\
136 & B & 21:30:47.3 & +50:49:03.1 & 0, 0 & & & 2.1(0.3) & 
0.46(0.05) & 0.37(0.07) & $<$0.4 \\
\underline {\bf 155} & B & 23:14:01.9 & +61:21:22.0 & +9, $-$5 & & & 1.7(0.2) & 
0.31(0.06) & 0.20(0.04) & 0.11(0.06) \\
160 & C & 23:38:30.1 & +60:53:43.0 & +8, +8 & & & 2.6(0.2) & 0.84(0.05) & 
0.52(0.03) & 0.20(0.07) \\ \hline
\multicolumn{11}{c}{Faint sources or sources without clear peak} \\ \hline
36 & B & 18:01:25.1 & $-$24:29:00.0 & & & & & 0.15(0.04) & & \\
57 & A & 18:25:37.8 & $-$07:42:19.9 & & & & & 0.13(0.03) & & \\
\underline {\bf 68} & A & 18:39:39.8 & $-$04:31:34.9 & & & & & 0.15(0.03) & & \\
87 & A & 18:58:38.1 & +01:06:57.0 & & & & & 0.06(0.04) & & \\
122 & A & 20:21:43.3 & +39:47:39.0 & & & & & 0.15(0.03) & & \\
125 & A & 20:27:51.0 & +35:21:33.0 & & & & & 0.13(0.05) & & \\
\underline {\bf 129} & A & 20:33:21.3 & +41:02:53.1 & & & & & 0.36(0.03)& & \\ 
\hline
\multicolumn{11}{c}{Undetected Sources} \\ \hline
66 & A & 18:36:23.1 & $-$05:54:58.9 & & & & & $<$0.1 & & \\
70 & A & 18:42:25.5 & $-$03:29:59.1 & & & & & $<$0.1 & & \\
86 & A & 18:57:10.6 & +03:49:22.0 & & & & & $<$0.2 & & \\
\underline {\bf 91} & A & 19:01:15.5 & +05:05:19.0 & & & & & $<$0.2 & & \\ 
\hline
\multicolumn{11}{c}{Missed Primary Peak} \\ \hline
\underline {\bf (3)$^a$}& S-D & 00:42:05.4 & +55:30:54.1 & +8, $-$4 &  & 0.6(0.5) 
& 0.16(0.02) & 0.017(0.007) & & \\
118 & A & 20:10:38.0 & +35:45:42.0 & & & & 1.0(0.2) & 0.25(0.05) & 
0.12(0.04) & $<$0.3 \\ \hline
\end{tabular}
\parbox{7in}{$^{\clubsuit}$ underlined and bold-face indicates
association with radio continuum emission (Paper~II).} \\
\parbox{7in}{$^{\rm a}$see footnote~\ref{newvla}.}\\ 
\parbox{7in}{$^{\rm b}$no FIVEPOINTS cycle done; however, we have
SCUBA maps (unpublished) confirming that the pointed position is
centered on a mm core.}\\ 
\parbox{7in}{$^{\rm c}$first code refers to 0.35-0.45mm photometry.}\\
\parbox{7in}{$^{\rm d}$0.45mm peak is 12\asec E, 2\asec S of 1.1mm peak.}\\
\end{flushleft}
\end{table*}

The results of the observations are listed in Table~\ref{obstab},
organized as follows; [Column~1] source number as in Paper~I; [Column~2]
code referring to the observing night (see Table~\ref{jou});
[Columns~3-4] source coordinates from the IRAS PSC-2; [Columns~5]
($\Delta \alpha$(\asec), $\Delta \delta$(\asec)) offset of 1.1 mm peak
from IRAS position of  the 1.1~mm emission peak to the IRAS coordinates;
[Columns~6-11] fluxes (in Jy)  observed at the given bands, with errors
(in Jy) in parentheses (including  both statistical and calibration
uncertainties); upper limits are given at the 3$\sigma$ level.

\begin{figure*}
\centerline{\hbox{
\psfig{figure=h1750f3.1,height=1.8in}
\psfig{figure=h1750f3.2,height=1.8in}
\psfig{figure=h1750f3.3,height=1.8in}
}}
\end{figure*}
\begin{figure*}
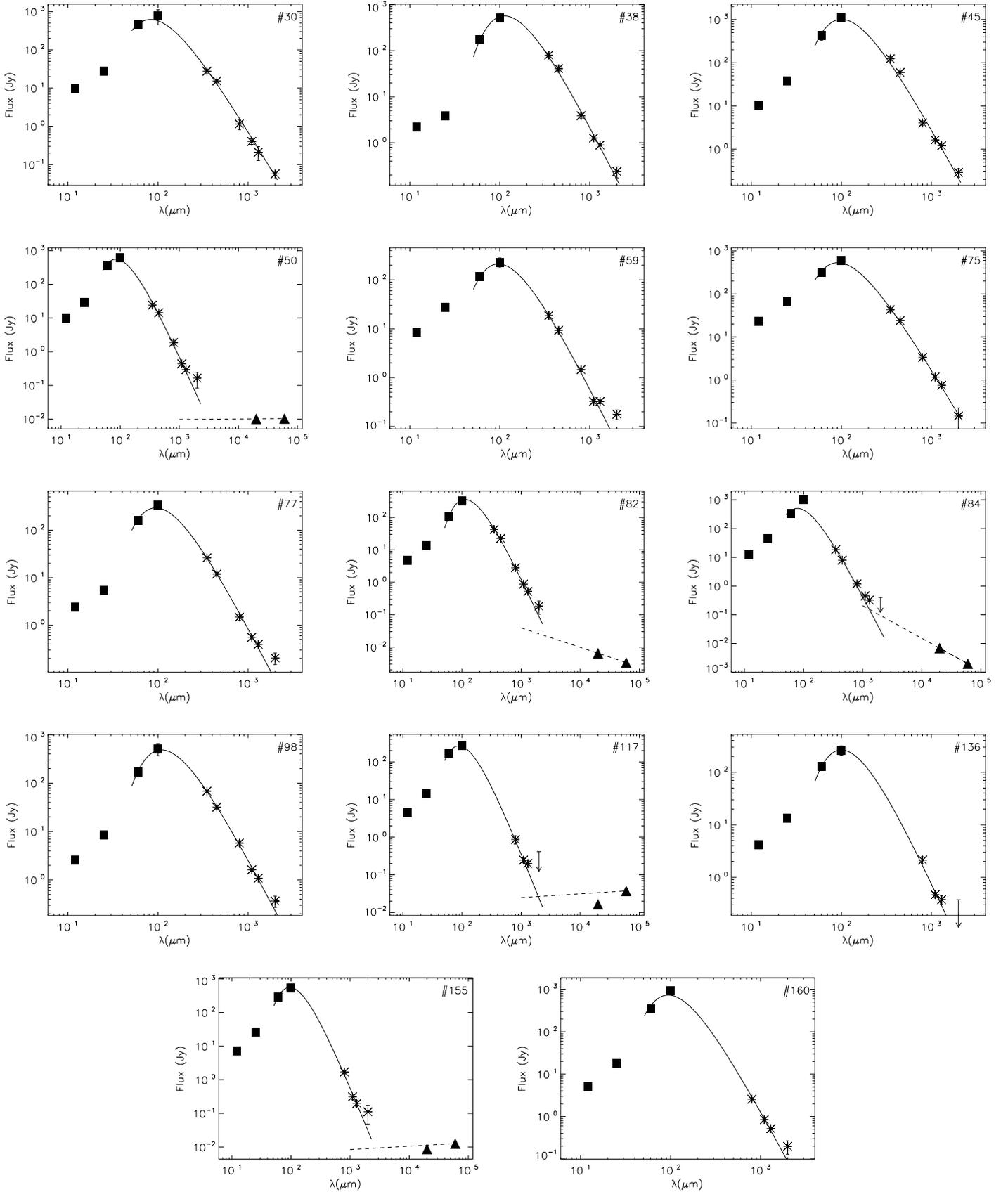

\centerline{\hbox{
\psfig{figure=h1750f3.4,height=1.8in}
\psfig{figure=h1750f3.5,height=1.8in}
\psfig{figure=h1750f3.6,height=1.8in}
}}
\centerline{\hbox{
\psfig{figure=h1750f3.7,height=1.8in}
\psfig{figure=h1750f3.8,height=1.8in}
\psfig{figure=h1750f3.9,height=1.8in}
}}
\centerline{\hbox{
\psfig{figure=h1750f3.10,height=1.8in}
\psfig{figure=h1750f3.11,height=1.8in}
\psfig{figure=h1750f3.12,height=1.8in}
}}
\centerline{\hbox{
\psfig{figure=h1750f3.13,height=1.8in}
\psfig{figure=h1750f3.14,height=1.8in}
\psfig{figure=h1750f3.15,height=1.8in}
}}
\centerline{\hbox{
\psfig{figure=h1750f3.16,height=1.8in}
\psfig{figure=h1750f3.17,height=1.8in}
}}
\caption{The spectral energy distribution of the 17 detected \lo\
sources (see Sect.~\ref{detections}) are reported below as plots of Flux
density (Jy) against wavelength (\um). IRAS photometric points are
represented by squares; triangles represent radio 2 and 6 cm fluxes when
a radio counterpart is present. Submillimeter photometry, after removal
of free-free contribution (see footnote~\ref{freefree}), is indicated
with asterisks.  The solid lines represent model fit to the data as
described in Sect.~\ref{sedmodel}; the corresponding parameters are
listed in Table~\ref{restab}. The SED for \#12 does  not report the
recently detected 3.6 cm emission.}
\label{sedplots}
\end{figure*}

Inspecting the spectral energy distributions plotted in
Fig.~\ref{sedplots}, one notes that in most cases the point at 2.0~mm,
and sometimes also the point at 1.3~mm, lie above the fitted curve 
(compare also with the fits obtained as described in
Sect.~\ref{sedmodel}); we believe this is due to the different beamsizes
of the UKT14 instrument at different wavelengths (see Sect.~\ref{obs}).
The ratios between areas of the beam at the different wavelengths are
A$_{1.3}$/A$_{(0.35-1.1)}$=1.1 and A$_{2.0}$/A$_{(0.35-1.1)}$=2.1; this
means that in case of a source uniformly filling the 2.0~mm beam
(27\asec), the measured 2.0~mm flux will be about a factor of 2 higher
than what an extrapolation from lower wavelengths might suggest. Scaling
1.3 and 2.0~mm photometry by these factors is in most cases sufficient
to bring those points to agree within the errorbars with the trend
suggested by the 0.35--1.1~mm fluxes. A proper scaling of the fluxes
would require knowledge of the (unknown) spatial brightness
distribution.

The luminosities of our sources as listed in Paper~I are based on the 4
IRAS-PSC2 fluxes, plus a correction to take into account emission
longward of 100\um\ (Cohen~\cite{C73}); this correction is likely to
overestimate the true luminosity, because it is assumed that the SED
falls off as a black body. With the help of the new millimeter and
submillimeter photometry we can derive a more accurate value for the
luminosity of the observed sources. 

With respect to Paper~I, the distances have also been slightly revised
for some sources. In Paper~I all distances were estimated from the
V$_{LSR}$ of the (1,1) ammonia line, using the observed velocity field
as determined by Brand \& Blitz (\cite{BB93}); however, as noted by
Brand \& Blitz, the data are quite sparse for distances greater than
5~kpc from the Sun in the II$^{nd}$ and III$^{rd}$ galactic quadrant. In
these cases one ought to estimate distances by using the analytical
expression for the galactic rotation curve, rather than the observed
velocity field; distances (and luminosities) for the affected sources 
have been updated accordingly.

\subsection{Association of IRAS \lo\ sources with millimeter
counterparts}
\label{statistics}

10 out of the 30 observed \lo\ sources are  associated with radio
continuum emission (Paper~II, but see footnote~\ref{newvla}).  We
consider a source associated with a millimeter continuum counterpart, 
and hence a detection, only when a {\it peak} of emission is clearly 
detected. In 17 of the 30 observed  sources we detected peaked
millimeter emission. In 7 (out  of 30) cases only faint millimeter
emission was detected,  or no clear peak was found; in 4 (out of  30)
cases only upper limits could be established.   In two cases (sources
\#3 and \#118) only a relative maximum of millimeter emission was
located by the automatic maximization procedure; these sources will be
conservatively considered as non-detections. These numbers are
summarized in  Table~\ref{detections}, where we also give information
about association with radio  counterpart (Paper~II) and H$_2$O masers
(Palla et al.~\cite{Petal91}).

The overall millimeter detection rate is 17/30$\sim$57\%; the mm
detection rate does not distinguish between \lo\ sources with
(6/10=60\%) and without (11/20=55\%) a radio counterpart.  Almost all
(9/10) sources with H$_2$O masers are associated with a mm peak, while
none of the sources with a maser has a radio counterpart. Because
the occurrence of water masers generally preceeds  the development of 
an UC\,\hii\ region (Churchwell et al.~\cite{CWC90}),
this result, together  with the 57\% overall detection rate, is an
independent confirmation of the validity of our approach to select
massive protostars (see Fig.~\ref{selection}).

In the group of \lo\ sources, we find young and old objects (see Sect.~1
and Paper~II), i.e. both UC\,\hii\ and older, extended \hii\ regions.
This way three out of four \lo\ sources with a radio counterpart that
are not associated with a mm-peak can be accounted for: in these sources
the associated \hii\ regions are more extended ($<$2, 225, 2500 and
3600\asec$^2$ for sources \#3 (see footnote~\ref{newvla}), 68, 91 and
129  respectively) than in 5 of the 6 \lo\ sources with both an \hii\
region and a peak in the mm-emission ($<$2, 100, 3.8, 3.7, and 
150\asec$^2$ for sources \# 12 (see footnote~\ref{newvla}), 50, 82, 84,
and 155S  respectively). Only \# 117 has relatively extended radio
emission (625\asec$^2$) and associated mm-emission. The lack of
millimeter detection correlates  with the extension of the \hii\ region
and identifies the oldest sources of the sample. For the \lo\ sources
without a mm counterpart, the ammonia column density never exceeds 
8$\times10^{13}$ \cmtwo\ (Paper~I), while it  ranges between 10$^{13}$
and 10$^{15}$ \cmtwo\ in the \lo\ sources detected in  the millimeter.
Hence one possible explanation for the non-detections in the millimeter
is that the peak of  emission is significantly displaced with respect to
the nominal IRAS position, and our ammonia observations only detected
the relatively lower-density peripheral regions of the core. Typically,
at least two or three FIVEPOINTS crosses (see Sect.~\ref{obs}) were
performed around the IRAS position, so that we could pick up the
millimeter peak only if it was within $\sim$20\asec\ from that
position.  Alternatively, the sources are not necessarily  compact; the
IRAS resolution at 100~\um\ is $\gsim$ 3\amin, and the bulk of the FIR
flux might arise from a relatively diffuse source. In this case either
the column density is too low for millimeter detection, or we might have
been chopping with both the ON and OFF beams in the diffuse source.  

\begin{table}
\begin{flushleft}
\caption{Detection Summary}
\label{detections}
\begin{tabular}{ccccc}\hline
mm & Radio & \lo$^a$ & H$_2$O maser$^b$ \\ \hline
Y & N & 11 & 8 \\
Y & Y & 6 & 0 \\
N & N & 9 & 1 \\
N & Y & 4 & 0 \\ \hline
\end{tabular} \\
\parbox{2in}{$^a$ radio detections include the two recent detections 
(see footnote~\ref{newvla}).} \\
\parbox{2in}{$^b$ based on Palla et al.~(\cite{Petal91})}\\
\end{flushleft}
\end{table}

\subsection{Dust physical  parameters}
\label{sedmodel}
It is common practice to use submillimeter and millimeter radiation to
trace the global properties of the dust. The assumption is that dust is
optically thin at submillimeter and millimeter wavelengths up to column
densities $N_{\rm H}\sim10^{25}$~\cmtwo\ (Mezger ~\cite{M94}). In the
usual formalism (Hildebrand \cite{H83}), the emission is assumed to come
from dust  at a single temperature and density. The flux observed at
each frequency can be expressed as:

\begin{equation}
F_{\nu}~=~{{M_{\rm D}}\over {d^2}}~\kappa_{\nu}~B(\nu, T)
\label{hilde}
\end{equation}

\noindent
where $M_{\rm D}$ is the dust mass, $d$ the distance, $\kappa_{\nu}$ the
dust mass opacity  parametrized as $\kappa_{\nu}=\kappa
_{\nu_0}(\nu/\nu_0)^{\beta}$, and $B(\nu, T)$ the Planck function. The
temperature enters only in the Planck function and  determines the
wavelength of the peak of the continuum emission. The mass affects the
overall level of continuum, while the opacity fixes both the absolute
level and the slope of the submm continuum. The largest uncertainty in
the mass determination comes from the assumption on the dust mass
opacity. Hildebrand (\cite{H83}) proposed a total {\it gas+dust} mass
opacity of 0.1 cm$^2$ g$^{-1}$ at 250~\um, and in spite of the
order-of-magnitude uncertainties that are generally believed to affect
the $\kappa_{250}$ value, Ossenkopf \& Henning (\cite{OH94}) concluded
that a variation at most of a factor 5  can be expected depending on the
presence of ice mantles on grains. To estimate the dust mass,
temperature and emissivity law, we fit Eq.(\ref{hilde}) to the available
data points by minimising the $\chi ^2$. For sources associated with
radio counterpart we extrapolated\footnote{For sources \#117 and 155,
the radio spectral indices (see Paper II) are not consistent with
free-free (either thin or thick) or ionised wind, but seem to suggest
non-thermal origin. We believe this is due to the extension of the
sources and the different $u-v$ coverage of our radio maps (Paper II)
which may result in loss of diffuse 2 cm emission. In this case we
extrapolated the 6 cm flux towards the sub(mm) assuming an optically
thin free-free spectrum.\label{freefree}} the observed 2 and 6 cm radio
continuum (Paper~II) and subtracted this contribution from the observed
millimeter fluxes before doing the fit. Twenty-five iterations were
performed in which the search radius for minimum $\chi ^2$ in each
variable was decreased by a factor 1.25 each time a minimum of $\chi^2$
was reached.  Reduced $\chi^2$ values at the end of the procedure were
typically less than 3-4.  We checked the repeatability of our results by
running the fitting procedure with starting values for mass, temperature
and dust opacity spanning one order of magnitude; the maximum variation
in the final best fit values was of less than 2\%, with the $\chi ^2$
remaining constant to the second decimal digit. We also checked the
sensitivity of the final $\chi ^2$ as a function of the fit parameters,
and we found that keeping the temperature fixed to a value within 10\%
of the best fit value yielded a $\chi ^2$ higher by 80\%, causing the
fit to converge at mass and $\beta$ values different by  respectively
$\sim$30\% and $\sim$10\% from the best fit values. The repeatability
and the high sensitivity of the $\chi ^2$ to the fit parameters, suggests that the internal accuracy of the method is within a few percent.

The fitted spectral energy distributions are presented in
Fig.~\ref{sedplots}. The peak of the continuum distribution  is around
100~\um\ so the submm data only cannot guarantee a meaningful
convergence of the fit. Therefore, we have also included the  IRAS 60
and 100~\um\ fluxes and for each source computed, based on the fitted
opacity and assuming an emitting area equal to the JCMT beamsize, the
wavelength where the optical depth becomes greater than unity (in most 
cases $\lambda<$60~\um). Having estimated the mass, we can then derive
the H$_2$ column density, assuming a gas/dust ratio of 100 by weight and
a size of the emitting area equal to the beam size.  The results of our
analysis for the 17 sources with a millimeter peak are summarized in 
Table~\ref{restab} organized as follows:  [Column~1] source running
number as in Papers~I and II; [Columns~2-3] distance to the source in kpc
and bolometric luminosity corrected as explained at the end of
Sect.~\ref{results}; [Column~4] $\beta$ value; [Column~5] total
(gas+dust) mass; [Column~6] derived H$_2$  column density; [Column~7]
dust temperature; [Column~8] wavelength  where $\tau$=1. 

\begin{table}
\begin{flushleft}
\caption{Derived Physical Parameters}
\label{restab}
\tabcolsep0.15cm
\begin{tabular}{crrccccc}\hline
\#$^{\dag}$ & \multicolumn{1}{c}{d} & \multicolumn{1}{c}{L} & $\beta$ & M$_{tot}$ 
& N(H$_2$) & T$_d$ & $\lambda _{\tau=1}$ \\
 & (kpc) & (\lsol) & & (\msol) & ({\tiny $10^{22}$cm$^{-2}$}) & (K) & (\um~) \\
 \hline
8   & 10.80 & 39300 & 1.56  & 210 & 1.7 & 37 & 12 \\
\underline {\bf (12)}$^a$ & 1.17 & 470 & 1.62  & 8.8 & 5.9 & 27 & 29 \\
28  & 4.48  & 5670  & 1.57  & 9.8 & 0.4 & 45 & 5 \\
30  & 2.00  & 3500  & 1.98  & 17.6 & 4.1 & 35 & 37 \\
38  & 0.12   & 6.4  & 2.08   & 0.34 & 21.8 & 24 & 91 \\
45  & 4.33  & 21200 & 1.89  & 360 & 17.7 & 29 & 71 \\
\underline {\bf 50} & 4.89 & 17300 & 1.98 & 105 & 4.1 & 34 & 37 \\
59  & 5.70  & 11000 & 1.77  & 100 & 2.8 & 32 & 23 \\
75  & 3.86  & 13000 & 1.58  & 100 & 6.2 & 35 & 27 \\
77  & 5.26  & 9000 & 1.72  & 115 & 3.8 & 32 & 26 \\
\underline {\bf 82}  & 6.77 & 15400 & 2.00 & 520 & 10.5 & 26 & 60 \\
\underline {\bf 84}  & 2.16  & 4300 & 2.03 & 12.2 & 2.4 & 37 & 30 \\
98  & 4.48  & 9200 & 1.65  & 270 & 12.4 & 29 & 47 \\
\underline {\bf 117} & 8.66  & 25100 & 2.04 & 190 & 2.5 & 33 & 30 \\
136 & 6.22  & 11600 & 1.87  & 200 & 4.8 & 30 & 35 \\
\underline {\bf 155} & 5.20  & 10600 & 2.38 & 200 & 6.8 & 28 & 65 \\
160 & 4.90  & 16000$^b$ & 2.03  & 230 & 8.8$^c$ & 31 & 130 \\
\hline
\end{tabular}\\
\parbox{3.25in}{
$^{\dag}$ Underlined and bold-face indicates association with radio 
counterpart.\\
$^a$ See footnote~\ref{newvla}.\\
$^b$ Luminosity from Molinari et al.~\cite{Metal98b}.\\
$^c$ Additional radio interferometry data indicate a Column density of 
$2\times 10^{24}$ \cmtwo. However, for homogeneity reasons we do not use 
this value in this table and in Fig.~\ref{lbol-ncol}.}
\end{flushleft}
\end{table}

\section{Global properties of the millimeter counterparts of \lo\
sources}
\label{global}
We will now turn our attention to the properties that can be derived
for  the mm-detected \lo\ sources. The exponent $\beta$ of the dust 
emissivity spans a range 1.56$\leq \beta \leq$2.38 with a mean  value of
1.86$\pm$0.23, consistent with the classical ``astronomical silicate''
(Draine \& Lee~\cite{DL84}) and with laboratory measurements (Agladze
et  al.~\cite{Aetal96}). For low-mass objects it has been found that older sources tend to have a higher $\beta$-value  (e.g. Zavagno et
al.~\cite{Zetal97}), a fact that can be explained by the destruction of
fluffy dust aggregates (Ossenkopf \& Henning~\cite{OH94}) due to the
increased envelope temperatures and higher energy  radiation fields
characteristic of more evolved YSOs. The difference in average $\beta$ 
values between \lo\ sources with and without associated radio emission, 
2.00$\pm$0.22 and 1.79$\pm$0.18 respectively, is only marginally 
significant and seems to suggest that among \lo\ sources detected in the 
millimeter, those which do not have a radio counterpart are younger.

The mean dust temperature T$_d$ is $32\pm 5$~K, consistent with the
shape of  the spectral energy distributions peaking at $\lambda
\sim$100~\um. The individual  values are up to 15~K higher than the
kinetic temperatures  derived from the ammonia observations (mean value
$24\pm 7$~K, excluding source \#75 whose temperature considerably
deviates from  the rest of the group).  This difference is in part due
to the different beam  sizes involved ($\sim$20\asec\ at the JCMT
against $\sim$40\asec\  at Effelsberg), but it may imply that millimeter
continuum observations trace denser and hotter material.

The total mass of circumstellar matter spans three orders of magnitude
and correlates well with the bolometric luminosity (both parameters have
the same dependence on distance). There is no difference in column
densities\footnote{We point out that we have limited information
about  the spatial distribution of the submillimeter emission in our
sources, so that the derived column densities could be lower limits in
case the source is smaller than the beam.} between \lo\ sources
associated and not associated with radio counterparts; the mean
value is $10^{22.7\pm 0.4}$ \cmtwo. Using the NH$_3$ column densities
from paper~I, we find an average ratio  6.3$\times 10^{-10}\leq$
[N(NH$_3$)/N(H$_2$)]$\leq 2.5\times 10^{-8}$. This value of the ammonia
abundance is consistent with the determinations of 3$\times 10^{-8}$ by
Harju et al. (\cite{Hetal93}), and 2$\times 10^{-9}$ and 9$\times
10^{-8}$ by Cesaroni \& Wilson (\cite{CW94}).
 
\section{The nature of the \lo\ sources with millimeter and without
radio counterparts: ZAMS stars or pre-ZAMS objects?}
\label{lownature}
We have seen that the group of \lo\ sources is somewhat of a mixed bag,
as it contains both young and old(er) objects. The youngest members of
this group are the \lo\ sources, detected at mm-wavelengths, but without
associated radio continuum emission. What we do not yet know, is the
actual evolutionary state of these latter objects: are they pre-ZAMS
objects, still  in the phase of mass-accretion (i.e. Class~0 objects),
or have they already reached the ZAMS, and therefore although young,
already in a more advanced evolutionary state? In the following
discussion we take a look at both  alternatives. In particular, we
must explain the lack of radio emission in our sample: such emission can
originate in an ionized stellar wind or in an \hii\ region. The latter
case is discussed below. Powerful, ionized winds are present both in the
pre-ZAMS and in the ZAMS phases. However, their emission generally is
much weaker than that of an UC\hii\ region and it would easily escape
detection given the typical distances of our sources (2--6 kpc).  For
example, the Orion BN object, a prototypical embedded, high luminosity
YSO, has a flux at 15 GHz of only 6.5 mJy (Felli et al.~\cite{Fetal93}),
too faint to be detected at distances greater than 2 kpc.

\subsection{They are ZAMS stars}
\label{zams}

The key question to answer in this case is why high-luminosity ZAMS
sources,  associated with massive circumstellar envelopes and
potentially able to create an \hii\ region, remain  undetected in radio
continuum. Various mechanisms that might be responsible will be looked
at. Two of the possible scenarios, namely the one invoking residual
accretion from the envelope to quench the formation of an \hii\ region,
and the possibility of a circumstellar disk origin for an ionised region
(Hollenbach et al.~\cite{Hetal94}),  have been discussed in Paper~II;
although they represent viable explanations, they will not be further 
elaborated here as our present data do not add new information on the 
subject.

\subsubsection{Compact and thick \hii\ regions}
\label{thickhii}
We start by examining the possibility that the sources are not detected
in the radio because the hypothetical \hii\ region is extremely compact
and optically thick. Our millimeter observations prove that these
sources are associated with peaked emission which implies average
particle densities of the order of 10$^5$~\cmthree\ at least; this
estimate is obtained dividing the column densities from
Table~\ref{restab} by the beam projected diameter at the various sources
distances, under the hypothesis of a spherical isodense and isothermal
dust envelope which completely fills the beam. This is a lower limit, however, since the sources might be smaller than the beam.
The central density of the YSO's envelope is the
parameter that mostly influences the expansion of an \hii\ region. The
volume of the initial Str\"omgren sphere of radius $r_{\rm S}$ is
inversely proportional to the 2/3 power of the density of the medium;
once the sphere is filled with hot ($\sim10^4$K) ionized material, the
pressure unbalance with the surrounding neutral and colder material
drives a rapid  expansion according to  (Spitzer~\cite{S78})

\begin{equation}
r(t)=r_{\rm S}\left(1+{{7c_{\rm s} t}\over {4r_{\rm S}}}\right)^{4/7},
\label{expansion}
\end{equation}

where $c_s$ is the isothermal sound speed and $t$ the time. It is
interesting to compare the radius reached by the \hii\ region with the
upper limits for the radii of optically thick \hii\ regions that we can
derive from our non-radio detections (Paper~II).  The main beam
brightness temperature of a homogeneous and opaque \hii\ region is given
by:

\begin{equation}
T_{\rm mb} = 10^{-26} {{F \lambda ^2}\over {2k\Omega}}
\end{equation}

where the flux $F$ is expressed in mJy, the wavelength $\lambda$ in
centimeters and the beam solid angle $\Omega$ in sterad.  A conservative
4$\sigma$ flux upper limit of 1~mJy at $\lambda=2$~cm in a 9.4$\times
10^{-11}$~sterad (or 2\farcs2) beam (Paper~II) for the radio emission in
our sources, translates into $T_{\rm mb}\sim$1.5~K. Assuming an electron
temperature of 10$^4$~K,  yields a beam filling factor of $T_{\rm
mb}$/10$^4$ = 1.5$\times 10^{-4}$, which allows us to express the upper
limit of the \hii\ radius  as a function of distance, $d$, as 

\begin{equation}
r_{{\rm ul}}~({\rm pc})\sim 6.7\times 10^{-8}\,d\,\,({\rm pc}).
\label{upplim}
\end{equation}

\begin{figure}
\centerline{\psfig{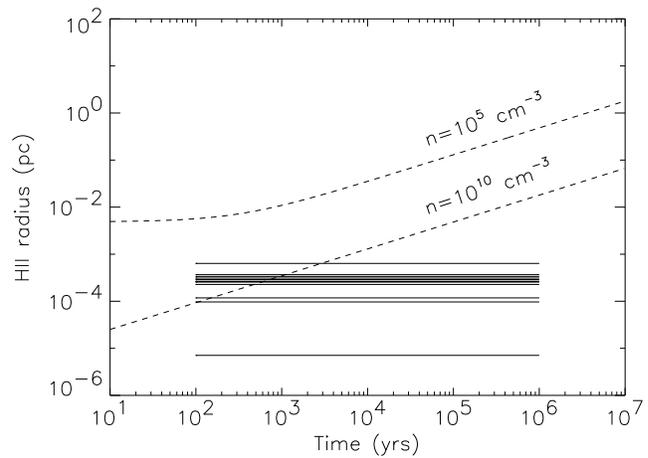}}
\caption[]{Plot of the radius $vs$ time for an expanding \hii\ region
around  a B0 ZAMS star for two values of the circumstellar density
(dashed lines). The horizontal lines mark the upper limit of the radius
of optically thick \hii\ regions computed with the parameters of our
radio non-detected sources.}
\label{hiiexpfig}
\end{figure}

The situation is then summarized in Fig.~\ref{hiiexpfig} (see also
DePree et al.~\cite{DPetal95}  and Akeson \& Carlstrom~\cite{AC96}). The
radius of the \hii\ region arising from a B0 ZAMS star (such a star has
a luminosity comparable to that of most of our sources - see
Table~\ref{restab}) is computed as a function of time according to
Eq.(\ref{expansion}) for two values of the circumstellar density, 10$^5$
and 10$^{10}$~\cmthree. From Fig.~\ref{hiiexpfig} it is clear that even
a circumstellar medium with a density of 10$^{10}$~\cmthree\ cannot keep
the size of the expanding \hii\ region below the estimated upper limits
for our sources for more than $\sim$3\,000 years. Therefore, it is
unlikely that our radio undetected sources with millimeter peak are
optically thick \hii\ regions, unless we accept that all of them are in
the first $\sim$3\,000 years of their expansion.  Comparing this time
with the estimated lifetime of an UC\,\hii\ region (Wood \&
Churchwell~\cite{WC89}), we should expect to find 100 times more \hii\
regions than precursors in our \lo\ sample; statistical arguments based
on the observations suggest a much lower number (see
Sect.~\ref{prezams}).

\subsubsection{Dusty \hii\ regions}
\label{dusthii}
We now consider the possibility that dust may completely absorb the UV
continuum from the central ZAMS star. Dust in \hii\ regions is needed to
explain why the Lyman continuum flux, derived from radio observations,
is generally lower than what is expected based on the bolometric
luminosity of the ZAMS star which drives the \hii\ region (e.g. Wood \&
Churchwell~\cite{WC89}; Paper~II). Dust grains which survive in the
ionized region have indeed the net effect to absorb a relevant fraction
of UV continuum emitted by the central star, which hence is no longer
available for ionization (Aannestad~\cite{A89}); the grain temperature
is a  function of optical properties, of the distance from the heating
source and the temperature of the heating source itself (in the present
case it is the stellar continuum of the newborn ZAMS star). The distance
where the grain temperature exceeds its sublimation value
($T_{\rm ddf}$=1500~K) sets the location of the dust destruction front
(``ddf'') and can be approximated as (Beckwith et al.~\cite{Betal90}):

\begin{equation}
R_{\rm ddf} = {{R_{\star}}\over {2}} \left({{T_{\star}} \over 
{T_{\rm ddf}}}\right)^{{4+\beta}\over {2}}
\label{ddf}
\end{equation}

This quantity should be compared with the size of an expanding \hii\
region computed from Eq.~(\ref{expansion}). In the case of a B0 ZAMS
star (T$_{\star}$=30\,900~K, R$_{\star}$= 5.5~\rsol;
Panagia~\cite{P73}), it can be shown that if the initial  circumstellar
density is $\leq10^8$~\cmthree\ then the dust destruction front is
enclosed in the ionized region already at the start of its expansion, 
irrespective of the value of $\beta$. In case of higher densities, the
ionized region is initially dust free, and the time needed for the \hii\
region to reach the ``ddf'' increases with density and $\beta$: about
3000~yrs are necessary for $\rho\sim 10^{12}$~\cmthree\ and $\beta \sim
2$. We note that in the short period when the expanding \hii\ region has
not yet reached the ``ddf'', it is its compactness which makes it
undetectable in the radio (see Sect.~\ref{thickhii}). In Paper~II  we
made the suggestion that the sources of the \lo\ sample that went
undetected in radio at the VLA, might be ZAMS stars with circumstellar
dust column densities high enough ($\ga 10^{22}$ \cmtwo) to completely
absorb the UV  field. Our millimeter continuum measurements allow us now
to check this  possibility. 

\begin{figure}
\centerline{\psfig{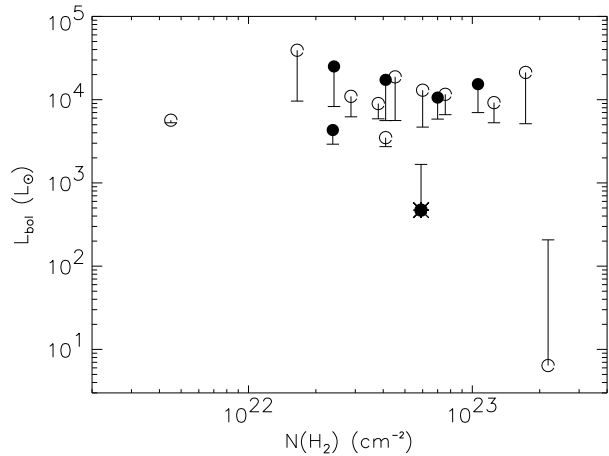}}
\caption[]{Bolometric luminosity plotted against H$_2$ column density
(from Table~\ref{restab}) for \lo\ sources associated with millimeter
counterpart. Full and empty circles represent \lo\ sources with and
without a radio counterpart, respectively. The small horizontal bars
connected to each symbol represent the luminosity threshold for radio
detection based on the parameters of our VLA observations, see Paper~II.
Source \#12 (indicated with a full circle and an asterisk) was 
recently detected in radio (see footnote~\ref{newvla}) but with a much
higher  integration time and in a different band and VLA configuration
with respect to  paper II.}
\label{lbol-ncol}
\end{figure}

In Fig.~\ref{lbol-ncol} we plot the bolometric luminosity against the
H$_2$ column densities given in Table~\ref{restab} for \lo\ sources with
millimeter counterpart both associated (full symbols) and not associated
(empty symbols) with a radio counterpart. In order to be sure that our
radio non-detections are not due to a luminosity effect, we perform on
each source the same check we did in Paper~II. Assuming optically thin
emission in the radio (the optically thick case has been treated in
par.~\ref{thickhii}.1) and that all sources are ZAMS stars, we convert
the 1~mJy upper limit (3$\sigma$) at 2~cm (Paper~II) into an upper limit
for the Lyman continuum flux  according to Eq.(5) of Paper~II, and into
an upper limit of the luminosity, using the stellar parameters of
Panagia (\cite{P73}). The luminosity thresholds for radio detection are
plotted in Fig.~\ref{lbol-ncol} as small horizontal bars connected by a
vertical segment to the corresponding value of the bolometric
luminosity. We see that in only two cases the expected luminosity is
greater than the observed one. This means that only for these sources
the stellar luminosity  is not high enough to produce a significant
amount of Lyman continuum photons.

Another important result is that a comparison between full and empty
symbols in Fig.~\ref{lbol-ncol} shows that both \lo\ sources associated
and not associated with radio counterpart span the same range in
luminosity and gas column density.  If the stellar UV continuum were
absorbed by the dust, we would expect the empty symbols to be preferably
found at high column densities, and the opposite for the full symbols.
The fact that we do not see such a segregation suggests that dust is not
responsible for the non-detection of radio continuum emission. 
However, a caveat is in order. If the size of the \hii\ region is much
smaller than that mapped by our submm continuum observations, no
correlation is expected between the column density given in
Table~\ref{restab} and that inside the \hii\ region. Then, dust
absorption  could account for the lack of radio emission. Observations
at submm wavelengths with arcsecond resolution should be valuable to
address this issue.

\subsubsection{Single sources or clusters ?}
\label{clusterhii}
Another possibility that could potentially explain the lack of radio
emission is that the IRAS sources contain a group or cluster of embedded
objects, a most likely occurrence for sources of spectral type earlier
than B5 (Hillenbrand~\cite{H95}, Testi et al.~\cite{Tetal99}). In such a
case, the total observed luminosity should be partitioned among all
cluster members with the result that the most massive object may no
longer  be bright enough to power a detectable \hii\ region.  To test
this hypothesis, we have computed the luminosity of the most massive
member assuming that masses in the cluster are distributed according to
the IMF (Miller \& Scalo~\cite{MS79}). We found (see also Wood \&
Churchwell~\cite{WC89}; Kurtz et al.~\cite{KCW94}; Cesaroni et
al.~\cite{Cetal94}) that the luminosity of the most massive member is
about 50\% of the total observed luminosity, and we can see from
Fig.~\ref{lbol-ncol} that for most of the sources a reduction of  the
luminosity to 50\% of the observed values would put them below their
individual detection thresholds (the horizontal dashes in
Fig.~\ref{lbol-ncol}). 

It is important to note however, that the cluster hypothesis works
equally well for \lo\ sources with and without radio counterpart. Given
that the two types of sources in our present sample have comparable
luminosities, the real question would be why sources are detected in
radio continuum. Thus, although it is likely that our sources contain a
cluster  of lower luminosity objects, we do not think that this 
occurrence can explain the lack of radio emission, unless the radio
detection identifies which sources are clusters and which are not.

\subsection{They are pre-ZAMS objects}
\label{prezams}

We now explore the possibility that our \lo\ sources with millimeter and
without radio counterpart are really precursors of UC\hii\ regions,
massive YSOs in a pre-ZAMS (pre-H-burning) phase, deriving their
luminosity from both accretion and contraction of their pre-stellar
core. 

\begin{figure}
\centerline{\psfig{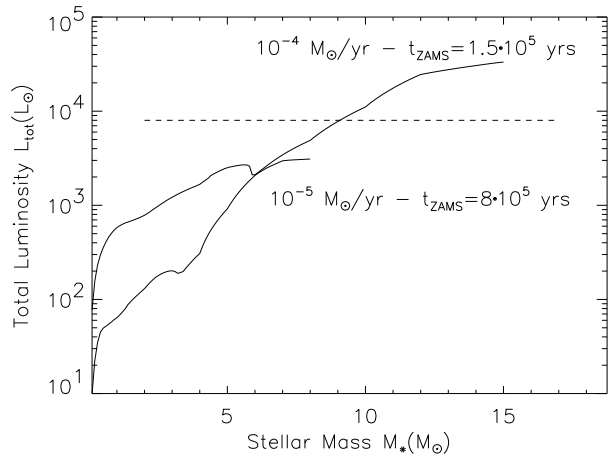}}
\caption[]{Total (accretion+surface) luminosity plotted against the
stellar core mass in solar units, for two values of the mass accretion
rate. The time needed to reach the ZAMS, is also indicated.}
\label{accr-fig}
\end{figure}

The pre-main sequence evolution of a massive object runs much faster
than for a low mass object. In particular, a $M_{\star}\geq $8\msol\
star accreting at a rate of 10$^{-5}$~\msunyr\ does not experience a
pre-main sequence phase, and the object joins the main sequence  while
still accreting mass from its parental cocoon (Palla \&
Stahler~\cite{PS90}). However, an 8 \msol\ mass star releases 
$\sim10^3$~\lsol\ on the ZAMS (Panagia~\cite{P73}), while it cannot
radiate more than $\sim3000$\lsol\ when accreting. The situation is illustrated in Fig.~\ref{accr-fig}, where the total
emitted luminosity is plotted against the core mass for mass accretion
rates of 10$^{-5}$ and 10$^{-4}$~\msunyr. In this simplified treatment
we assume that the emitted luminosity comes from accretion and the
contraction of the central pre-stellar core\footnote{We have used
$L_{acc}$=3.14$\times 10^4$\lsol ($M_{\star}$/\msol)(\rsol /$R_{\star}$)
($\dot{\rm M}/10^{-3}$\msol\,yr$^{-1}$), the $M_{\star}/R_{\star}$
relationship of Palla \& Stahler (\cite{PS91}), and the radiative
luminosity from Fig.~1 of Palla \& Stahler~(\cite{PS93}).}. 

Since most of our sources have luminosities in excess of  $\sim
10^4$\lsol, we conclude that we are either exploring a higher mass range
or higher rates of mass accretion. It is important to note that in the
latter case the mass at which the central object reaches the ZAMS
increases from 10~\msol\ for $\mdot = 3\times 10^{-5}$~\msunyr\ to
15~\msol\ for $\mdot = 10^{-4}$~\msunyr\ (Palla \& Stahler~\cite{PS92}).
In other words higher accretion rates will produce higher mass stars
when the star first reaches the ZAMS.

As shown in Fig.~\ref{accr-fig}, the end point of each curve corresponds
to the mass at the time of the arrival on the ZAMS, t$_{\rm ZAMS}$. We
can see that an object accreting at 10$^{-4}$\msunyr\ has a luminosity
comparable to those of our sources (say $\geq$ 8\,000\lsol, the
horizontal dashed line)   for $\sim$60\,000 years. On the other hand, a
protostar accreting at $\mdot= 10^{-5}$ \msunyr\  will never be able to
produce the observed luminosities. There are 8 sources with L$>$8000
\lsol in our sample, and this luminosity would  correspond to a core
mass of 9 \msol\ for an accretion rate of 10$^{-4}$\msunyr. A comparison
with the time that a massive source spends in the UC\hii\ phase
(t$_{\hii}\sim3\times 10^5$ yrs, Wood \& Churchwell~\cite{WC89})
suggests that our sample should contain $\sim$5 times more UC\hii\
regions than \hii\ precursors.  How does this number compare with the
observations?

The number of \lo\ sources with radio counterparts is 11 (also including
\#3 and \#12 -- see  footnote~\ref{newvla}), out of an original sample
of 37 \lo\ sources associated with ammonia (Paper I) and observed in
radio (Paper II). It is plausible to assume, however, that a number of
older \lo\ sources not associated with ammonia may be associated with
radio continuum emission. Indeed, as we pointed out in Paper~I, a 6 cm
VLA survey (Hughes \& MacLeod~\cite{HML94}) made on a list of sources 
containing 15 \lo\ from our original sample, showed that although the
radio detection rate was very high ($\sim$95\%), ammonia was absent in
80\% of the \lo\ sources detected in radio by Hughes \& MacLeod. This
suggests that 80\% of the \hii\ regions present in the \lo\ group are
not associated with ammonia. The total number of \hii\ regions we
estimate for the complete sample of \lo\ sources in Paper~I is then
11/0.2=55 sources.

Now, we estimate the number of UC\hii\ precursors. There are  8 \lo\
sources with L$\geq$8\,000\lsol\  associated with a peak of millimeter
emission without radio counterpart, corresponding to 40\% of the total
sample of 20 objects. Since not all \lo\ sources without radio
counterpart could be observed in  the submillimeter, the total number of
\hii\ precursors present in the  \lo\ group is estimated as
(38--11)$\times$0.4$\sim$11 (assuming that sources without ammonia
association do not contain \hii\ precursors).  

In conclusion, the ratio between  the number of \hii\ regions and \hii\
precursors is $\sim$5, remarkably close to the expected number. Also,
the number of \hii\ precursors agrees with the expectation of $\sim$10
sources in the entire \lo\ group given in Paper I.

\section{Conclusions}
\label{conclusions}

We have obtained millimeter and submillimeter photometry of a sample of
30 luminous  \lo\ sources known to be associated with dense gas.  The
aim of these observations is to identify those objects that might be
considered precursors of UC\,\hii\ regions.  A clear millimeter
continuum peak is detected in 11 sources out of 20 sources without a
radio counterpart, and in 6 out of 10 sources with a radio counterpart.
The derived dust temperatures and column densities do not distinguish
between the two types of sources. The mean value of $\beta$, the
exponent of the frequency dependence of the opacity, seems higher for
sources detected in the radio continuum (2.00$\pm$0.22 versus
1.79$\pm$0.18) which we interpret as an indication of a more advanced
evolutionary state.

As to the nature of the \lo\ sources detected in the (sub)mm, but
without  associated radio continuum emission, two alternative
explanations have been considered: they are either  pre-ZAMS objects
without Lyman continuum emission, or stars deriving their luminosity
from H-burning on the ZAMS. In the latter case, one has to justify the
lack of radio continuum  emission, and several possible explanations
have been examined: (i) compact, optically thick \hii\ regions; (ii)
total absorption of the ionizing flux by dust; (iii) presence of a
cluster of objects instead of a single central source; (iv) residual
accretion from an infalling envelope. In particular, we have shown that:

\begin{itemize}
\item[(i)] Based on statistical arguments, the number of \hii\
regions in our original sample of \lo\ sources exceeds the number of
\hii\ precursors by a factor 5. If the candidate \hii\ precursors were
optically thick UC\,\hii\ regions, this factor should be of the order of
100, too large to be acceptable. 

\item[(ii)] The possibility that dust may completely absorb the ionizing
UV  continuum seems also to be excluded: at comparable luminosities,
\lo\ sources not associated with radio emission do not have higher dust
column densities. However, an inverse proportionality seems to exist
between the dust column density and the extension of the \hii\ region,
when present.

\item[(iii)] The presence of a cluster of lower mass stars cannot be
excluded, but does not explain the fact that sources of comparable
luminosity do show radio emission.

\item[(iv)] The most likely explanation invokes residual accretion from
the  envelope onto a massive star/disk system. As discussed in Paper~II,
an accretion rate as low as $3-4\times 10^{-6}$~\msunyr\ should be able
to prevent the formation of an \hii\ region around a B0 star. The possibility that an \hii\ region is produced by the ionization of a circumstellar disk, also discussed in Paper~II, is also a viable explanation.
\end{itemize}

Considering the alternative possibility that the sources are pre-ZAMS 
objects, we have found that for an accretion rate of  $\sim
10^{-4}$~\msunyr, the expected ratio of \hii\ regions to \hii\
precursors should be $\sim$5, in excellent agreement with the value
estimated from the observations. In such case, these pre-ZAMS
objects should be characterised by higher accretion rates than  ZAMS
stars of the same luminosity.

In order to distinguish between the two plausible explanations (massive
protostars with fast accretion rates vs young ZAMS star with modest
residual accretion), high angular resolution millimeter observations are
being collected on the most promising candidates selected from the
present study. 

\begin{acknowledgements}
S.M. thanks G.G.C. Palumbo for the financial support provided for the
trip to the JCMT.  We acknowledge L. Testi for a critical reading of the
manuscript.  We are grateful to the James Clerk Maxwell Telescope staff,
and in particular G. Sandell, for their assistance during the
observations; the UKSERV program for remote service observations with
the JCMT is also acknowledged. The James Clerk Maxwell Telescope is
operated by The Royal Observatories on behalf of the UK PPARC, the
Canadian NRC and the Netherlands NWO. This project was partially
supported by ASI grant ARS-98-116. 
\end{acknowledgements}

\end{document}